\documentclass[twocolumn,floatfix,prl,aps]{revtex4-1}
\usepackage{graphicx}
\usepackage{amsmath}
\usepackage{amssymb}
\usepackage{color}
\usepackage{natbib}
\usepackage[pdftex,breaklinks,bookmarks=false,colorlinks,linkcolor=blue,citecolor=blue,urlcolor=blue]{hyperref}

\newcommand{\cev}[1]{\reflectbox{\ensuremath{\vec{\reflectbox{\ensuremath{#1}}}}}}

\newcommand{\ket}[1]{ | #1 \rangle }
\newcommand{\be}{\begin{equation}}
\newcommand{\ee}{\end{equation}}
\newcommand{\beq}{\begin{eqnarray}}
\newcommand{\eeq}{\end{eqnarray}}

\newcommand{\bz}{{\bf z}}

\newcommand{\makebf}[1]{\boldsymbol{#1} }

\begin{document}

\title{Fractional Excitonic Insulator}

\author{Yichen Hu, J\"orn W. F. Venderbos and C. L. Kane}
\affiliation{ Department of Physics and Astronomy, University of Pennsylvania, Philadelphia, PA 19104-6323, USA}

\begin{abstract}
We argue that a correlated fluid of electrons and holes can exhibit a fractional quantum Hall effect at zero magnetic field analogous to the Laughlin state at filling $1/m$.   We introduce a variant of the Laughlin wavefunction for electrons and holes and show that for $m=1$ it is the exact ground state of a free fermion model that describes $p_x + i p_y$ excitonic pairing.   For $m>1$ we develop a simple composite fermion mean field theory, and we present evidence that our wavefunction correctly describes this phase.   We derive an interacting Hamiltonian for which our wavefunction is the exact ground state, and we present physical arguments that the $m=3$ state can be realized in a system in which energy bands with angular momentum that differ by $3$ cross at the Fermi energy.   This leads to a gapless state with $(p_x + i p_y)^3$ excitonic pairing, which we argue is conducive to forming the fractional excitonic insulator in the presence of interactions.   Prospects for numerics on model systems and band structure engineering to realize this phase in real materials are discussed.
\end{abstract}

\maketitle

The quantum Hall effect was originally understood as a consequence of the emergence of Landau levels for two dimensional electrons in a magnetic field~\cite{klitzing1980}, but was reformulated in the framework of topological band theory~\cite{tknn1982}.   This introduced the notion of ``Chern bands", which have a rich structure due to the interplay between lattice translations and magnetic translations~\cite{hofstadter1976}, and allow for the existence of a Chern insulator in the absence of a uniform magnetic field~\cite{haldane1988}.  There is a sense in which all quantum Hall states are the same and can be adiabatically connected to a flat band limit that resembles a Landau level.   However, the {\it opposite} to the flat band limit occurs near a quantum Hall transition, which occurs when the conduction band and valence band invert at a Dirac point~\cite{ludwig1994}.   A weakly inverted quantum Hall state differs from a trivial insulator only near the Dirac point, and can be viewed as a quantum fluid formed by the low energy electrons and holes of the original trivial insulator.   The band inversion paradigm has proven to be a powerful tool for engineering topological phases of non-interacting fermions~\cite{readgreen2000,bhz2006,fukane2007,yu2010}.

In recent years there has been effort to study analogs of the Chern insulator for the fractional quantum Hall (FQH) effect.   Theoretical work has focused on the proposal for creating nearly flat Chern bands \cite{tang2011,neupert2011,sun2011} that can be fractionally filled and can host states---called fractional Chern insulators \cite{regnault2011}---that resemble the Laughlin state of a fractionally filled Landau level (see the reviews \cite{parameswaran2013, neupert2015, liu2013review} and references therein).   Experimental progress has been reported in twisted bilayer graphene~\cite{spanton2018}, where the commensuration with the moir\'e pattern leads to interesting structure in the observed FQH states at finite magnetic field.  The zero field fractional Chern insulator is more challenging because it requires a non-stoichiometric band filling.   Here we consider the opposite limit and propose a wavefunction describing a {\it fractional excitonic insulator}: a gapped FQH state built from a strongly correlated fluid of electrons and holes.   We argue that this provides an alternative route to realizing a FQH state at zero field in a stoichiometric system that is close to a special kind of band inversion. 

We consider a  wavefunction inspired by the celebrated Laughlin wavefunction~\cite{laughlin1983} of the form 
\begin{equation}
|\Psi_m\rangle = \sum_N \frac{f^N}{N!} |\psi^N_m \rangle,  \label{psin}
\end{equation}
 where $|\psi^N_m\rangle$ describes a state with $N$ electrons and holes described by a Jastrow wavefunction 
\begin{equation}
\psi^N_m(\{z_i,w_j\}) = \frac{\prod_{i<i'} (z_i - z_{i'})^m \prod_{j<j'} (w_j - w_{j'})^m}{\prod_{i,j} (z_i-w_j)^m}.
\label{wf}
\end{equation}
Here $z_{1, ..., N}$ ($w_{1, ..., N}$) are complex coordinates for electrons (holes) and $m$ is an odd integer.  $\psi^N_m$ is similar to a Halperin bilayer wavefunction~\cite{halperin1983}, except that the Gaussian associated with the lowest Landau level is absent, and it has a singular denominator.   The denominator can be fixed without changing the long distance behavior by introducing a cutoff $\xi$ in a prefactor $\prod_{ij} h(|z_i-w_j|/\xi)$, where $h(x\rightarrow 0)\sim x^{2m}$ and $h(x\rightarrow\infty) = 1$~\cite{kim2001}.   A similar wavefunction was mentioned by Dubail and Read~\cite{dubail2015} in connection with tensor network trial states.    Like them, we will argue that $|\Psi_m\rangle$ is topologically equivalent to a single component $\nu=1/m$ Laughlin state.

We will begin by showing that for $m=1$, $|\Psi_1\rangle$ (despite the denominator) is the {\it exact} ground state of a simple non-interacting model of a Chern insulator, and can be viewed as a condensate of $p+ip$ excitons.   We then present several pieces of evidence that $|\Psi_{m>1}\rangle$ describes a FQH state.   This includes an analysis of the Laughlin plasma analogy, as well as the ground state degeneracy on a torus.  We introduce a composite fermion mean field theory as well as a coupled wire model that reproduce the phenomenology of the FQH state.  We also identify an interacting Hamiltonian whose exact ground state is \eqref{wf}.     Finally, we propose that a feasible route towards realizing this state is to find a material whose band structure features the touching of two bands that differ in angular momentum by $3$.   We argue that coupling the bands favors excitonic pairing in a $(p_x+ip_y)^3$ channel, and that interactions could stabilize the $m=3$ state.

To describe the $m=1$ state, consider the non-interacting spinless fermion Hamiltonian,
\begin{equation}
{\cal H}_1 = \sum_{\bf k} \epsilon_{\bf k}(c_{e{\bf k}}^\dagger c_{e{\bf k}} +  c_{h{\bf k}}^\dagger c_{h{\bf k}}) + \Delta_{\bf k} c^\dagger_{e{\bf k}}c^\dagger_{h-{\bf k}} 
+ h.c.,\label{h2band}
\end{equation}
with
\begin{equation}
\epsilon_{\bf k} = (k^2 - v^2)/2; \quad \Delta_{\bf k} = i v(k_x - i k_y).
\label{hm=1}
\end{equation}
This is a two band model in which $c^\dagger_{e(h){\bf k}}$ create conduction band electrons (valence band holes).
We particle-hole transformed the valence band, so that the vacuum $|0\rangle$ (annihilated by $c_{e,h{\bf k}}$) is the topologically trivial filled valence band.   This model is properly regularized for $k\rightarrow\infty$, and describes a Chern insulator in which the conduction and valence bands are inverted at ${\bf k}=0$.  
Note that \eqref{hm=1} has a single parameter $v$
~\footnote{The phase of $\Delta_{\bf k}$ can be chosen by defining the phase of $c_{h{\bf k}}$.  The choice in (\ref{hm=1}) makes $f$ in (\ref{psin}) real.}.  
The coefficient of $k^2$ can be fixed by a choice of units, but a more generic model~\cite{liu2010,liu2011} has independent coefficients for the other terms.   For this particular choice the energy eigenvalues are $\pm E_{\bf k} = \pm (k^2 + v^2)/2$.   The analysis of this model is similar to the BCS theory of superconductivity.  The ground state is
\begin{equation}
|\Phi_{m=1} \rangle = \prod_{\bf k} (u_{\bf k} + v_{\bf k}c^\dagger_{e{\bf k}}c^\dagger_{h-{\bf k}})|0\rangle,
\end{equation}
where $u_{\bf k} = i(k_x+ik_y)/\sqrt{2E_{\bf k}}$ and $v_{\bf k}=v/\sqrt{2E_{\bf k}}$.
Following the Read Green analysis of a $p+ip$ superconductor~\cite{readgreen2000}, this can be written in the real space form
\begin{equation}
|\Phi_{m=1} \rangle \propto e^{\int d^2{\bf z} d^2{\bf w} g({\bf z}-{\bf w}) \psi^\dagger_e({\bf z}) \psi^\dagger_h({\bf w})}|0\rangle,
\label{psi g(z-w)}
\end{equation}
where $c^\dagger_{e,h{\bf k}}$ and $g_{\bf k}\equiv v_{\bf k}/u_{\bf k} = -iv/(k_x+ik_y)$ have Fourier transforms $\psi^\dagger_{e,h}(z=x+iy)$ and $g(z) = v/(2\pi z)$.   $|\Phi_{m=1}\rangle$ then has the form (\ref{psin}) with $f = v/(2\pi)$ and
\begin{equation}
\phi^N_{m=1}(\{z_i,w_j\}) = {\rm det}\left[\frac{1}{z_i-w_j}\right].   \label{psidet}
\end{equation}
The equivalence of $\phi^N_{m=1}$ and $\psi^N_{m=1}$ follows from the Cauchy determinant identity~\cite{milovanovic1995}, which can be checked by writing the determinant over a common denominator, noting its units and antisymmetry.

Though the precise form of $g({\bf z})$ that makes the Jastrow form exact is particular to our choice of parameters, the topological structure of the Chern insulator dictates that the $1/z$ behavior for $z\rightarrow \infty$ remains in a more generic theory.   The short distance behavior, however, depends on the details as well as the lattice cutoff.   A related model was studied in Ref. \onlinecite{liu2011}, where the connection was made to a Halperin $(1,1,-1)$ bilayer state.  Viewed as a bilayer system, this is related to a $(1,1,1)$ state by a particle-hole transformation in one layer~\cite{yang2001}.   The $(1,1,1)$ state describes a single component ``spin polarized" quantum Hall fluid with broken spin symmetry.   In our problem the spin symmetry corresponds to the independent conservation of electrons and holes, which is violated by the ``$p+ip$ pairing term" $\Delta_{\bf k}$.   Thus, we can view the Chern insulator as an {\it excitonic insulator} that is distinguished from the trivial insulator by a condensation of $p+ip$ excitons.  Unlike the original excitonic insulator~\cite{jerome1967,halperin1968}, this condensation does not involve a spontaneously broken symmetry, since electrons and holes are not independently conserved.   It is analogous to a {\it proximitized} $p+ip$ superconductor.

Encouraged by the success of $|\Psi_{m=1}\rangle$,
we now consider the generalization to a fractional excitonic insulator.   To motivate that this should be possible, we first introduce a composite fermion mean field theory.   Consider a $2D$ two band system and perform a statistical gauge transformation that attaches $\pm (m-1)$ flux quanta to the electrons (holes)
\cite{hlr1993}.   This is accomplished in Eqs. \eqref{h2band} and \eqref{hm=1} by replacing ${\bf k} c_{e(h){\bf k}} \rightarrow (-i \nabla \pm  {\bf a})\psi_{e(h)}$, where  the statistical vector potential satisfies
\begin{equation}
\nabla \times {\bf a} = 2\pi (m-1) (\psi_e^\dagger \psi_e - \psi^\dagger_h \psi_h).
\label{avec}
\end{equation}
Equivalently, in  a Lagrangian formulation, flux attachment is implemented by adding a Chern-Simons term ${\cal L}_{CS}= \epsilon_{\mu\nu\lambda} a_\mu \partial_\nu a_\lambda/(4\pi (m-1)) \equiv a\partial a/4\pi (m-1)$~\cite{zhk1989}.
This is different from the conventional composite fermion model, because in the valence band  flux is attached to the {\it holes} rather than the electrons.   This transformation has no effect on electrons deep in the valence band and is compatible with exact particle-hole symmetry \cite{son2015}.   

When the electron and hole densities are equal, the average statistical flux seen by each particle is zero.   Thus, in mean field theory we can consider a system of composite fermions with Hamiltonian given by \eqref{h2band} and \eqref{hm=1}.  Assuming the composite fermions are in a Chern insulator phase, we integrate them out in the presence of ${\bf a}$ and the external vector potential ${\bf A}$. This leads to ${\cal L}_{\rm eff} = {\cal L}_{CS} + (a + A)\partial (a+A)/4\pi$.  Integrating out ${\bf a}$ then gives ${\cal L}_{\rm eff} = A\partial A/4\pi m$.  This shows the resulting phase is a FQH state with $\sigma_{xy} = (1/m)e^2/h$. A second indication this phase is possible is provided by the coupled wire construction~\cite{kml2002}.   In Supplemental Section I~
\footnote{See Supplemental Material.} 
we show that an array of alternating $n$-type and $p$-type wires can support this phase at zero magnetic field.

\begin{figure}
\includegraphics[width=\columnwidth]{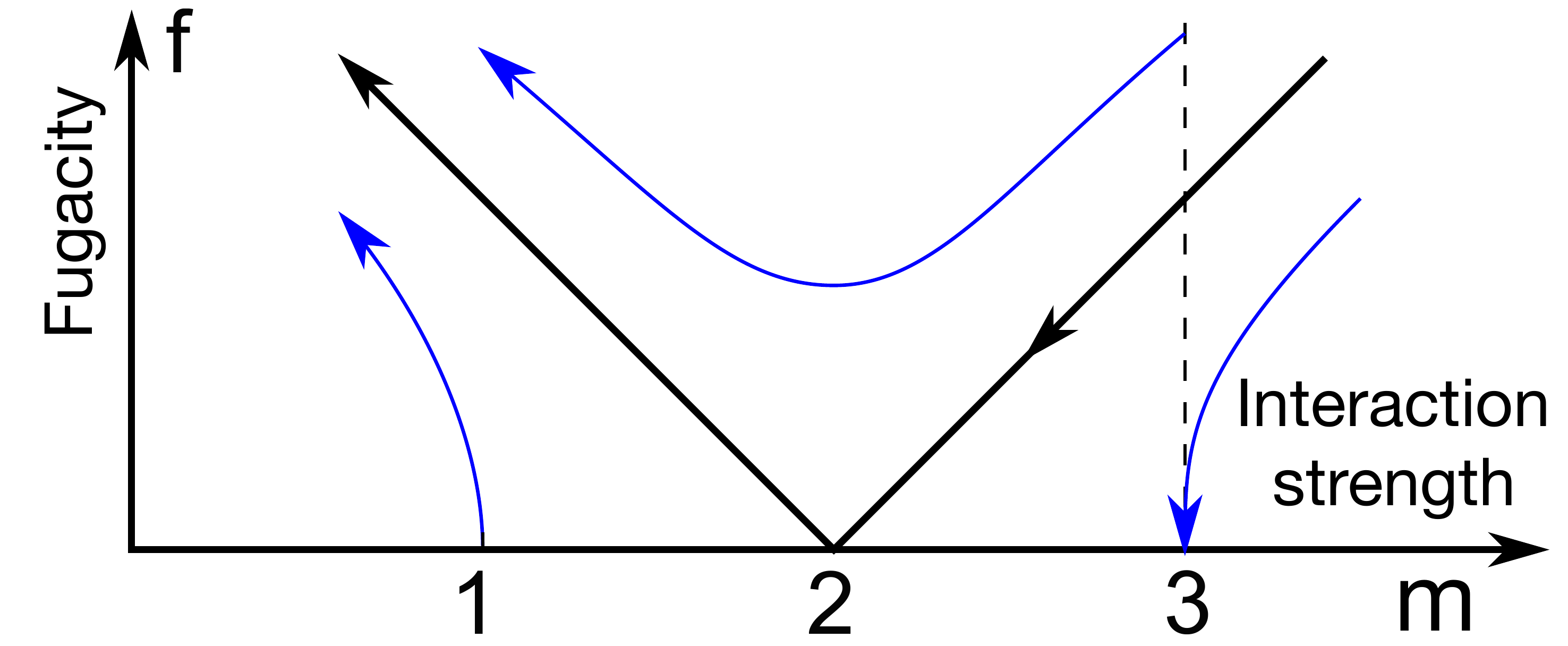}
\caption{Kosterlitz Thouless renormalization group flow diagram~\cite{kt1973} for the plasma analogy of \eqref{psin} and \eqref{wf} as a function of fugacity $f$ and the coefficient of the Coulomb interaction, the bare value of which is controlled by $m$.    }
\label{Fig1}
\end{figure}

We now analyze the wavefunction of Eq.~\eqref{psin} and \eqref{wf}.  To determine whether it describes a FQH fluid, we follow Laughlin~\cite{laughlin1983} and view $\langle \Psi_m|\Psi_m\rangle$ as the partition function of a classical plasma.   Like Laughlin's plasma, our charges interact by a 2D Coulomb interaction $-\beta V = \sum_{i<j} 2 m q_i q_j \log |z_i-z_j|/\xi$, where $m$ plays the role of inverse temperature.   Unlike Laughlin's plasma, our plasma has charges $q_i = \pm 1$, and the neutralizing background (due to the Gaussian) is absent.  It is in the grand canonical ensemble with a fugacity $f$.   This plasma maps precisely to the Kosterlitz Thouless problem~\cite{kt1973,dubail2015}, and exhibits two phases: a high temperature phase characterized by perfect screening, and a low temperature phase with bound charges.  For small $f$ the transition is determined by balancing the energy $m \log L$ of an unbound charge  with the entropy $\log L^2$ giving a critical point at $m=2$.  For $m=1$ the plasma is in the screening phase, which is consistent with our understanding of $|\Psi_1\rangle$ as a quantum Hall state.   For $m=3$ the plasma is in a bound phase for small $f$.   This is similar to the Laughlin wavefunction for large $m$, which describes a crystal.  However, for larger $f$ screening renormalizes the Coulomb interaction, and a screening phase is expected above a critical value of $f$, as indicated in Fig.~\ref{Fig1}.  Since the only length in the problem is the cutoff scale $\xi$, the screening phase will occur at high density, when electrons and holes have a typical separation of order $\xi$.

The structure of the plasma analogy is reminiscent of the wire construction for the $\nu=1/m$ state~\cite{kml2002}, which involves coupling edge states with an irrelevant sine-Gordon type coupling that leads to exactly the same plasma~\cite{Note2}.  The correspondence of the plasmas is not an accident, given the expectation that the ground state wavefunction can be interpreted as a correlator of the same conformal field theory that describes the edge states~\cite{mooreread1991}.   The only difference with the conventional Laughlin state is the absence of the background charge.   Following this logic, we construct a wavefunction for a quasi-hole at position $Z$ as
\begin{equation}
\psi^{e^*}_N(Z,\{z_i,w_j\}) = \prod_i \frac{Z-z_i}{Z-w_i} \psi_N(\{z_i,w_j\}).
\end{equation}
In the plasma analogy, this state has an external charge at $Z$.   Assuming the plasma perfectly screens, this leads to a charge $e^*=e/m$ quasi-hole.   Quasi-electron states are constructed similarly by exchanging $z_i$ and $w_j$.  

Another probe of  topological order is the ground state on a torus, which may also be useful for numerical studies.  Following Haldane and Rezayi~\cite{haldane1984}, we consider a torus with $z=z+L$ and $z=z + L \tau$ identified ($\tau$ is a complex number describing the shape of the torus).   The periodic generalization of (\ref{wf}) then involves two modifications.   First, the terms in the denominator become
\begin{equation}
(z_i-w_j)^m \rightarrow \vartheta_1( \pi(z_i-w_j)/L |\tau)^m,
\end{equation}
 where  $\vartheta_1(u|\tau)$ is the odd elliptic theta function~\cite{gradshteyn1980}.   The terms in the numerator are modified similarly.   Second, $\psi^N_m$ is multiplied by a function of the center of mass coordinates  $Z=\sum_i z_i$, $W=\sum_j w_j$, given by 
\begin{equation}
F_{CM}(Z,W) = e^{iK(Z-W)} \vartheta_1(\pi(Z-W-z_0)/L|\tau)^m .
\end{equation}
From the periodicity properties of $\vartheta_1(u|\tau)$, it can be checked that this modified wavefunction is properly periodic, with $K$ and $z_0$ depending on the phase twisted boundary conditions.   For fixed boundary conditions there are $m$ independent choices for $K$ and $z_0$, establishing the $m$-fold ground state degeneracy.   
We have also checked that for $m=1$ the non-interacting ground state of $(\ref{hm=1})$ on a torus has the form $\det[g(z_i-w_j)]$, with $g(z) \propto e^{i K z}\vartheta_1(\pi(z-z_0)/L|\tau)/\vartheta_1(\pi z/L|\tau)$.  ($K,z_0$ again depend on boundary conditions).   A generalization of the Cauchy identity~\cite{verlinde1986} shows that this is precisely equivalent to the wavefunction described above.

Having established that \eqref{psin} and \eqref{wf} describe an excitonic fractional quantum Hall state, we now seek a Hamiltonian that can realize it.   One approach is to find an ``exact question to the answer": a Hamiltonian designed to have $|\Psi_m\rangle$ as its exact ground state~\cite{arovas1992}.   While we do not have an analog of the two body $\delta$-function type interaction~\cite{trugman1985} that stabilizes the Laughlin state, we adopt the construction in Ref. \onlinecite{kane1991}, which provides a natural generalization of (\ref{hm=1}) to $m>1$ at the price of introducing several-body interactions.   By applying $\partial_{z_j^*} \equiv \frac12(\partial_{x_j} + i\partial_{y_j})$ (or $\partial/\partial_{w_j^*}$) to (\ref{wf}) and noting that due to analyticity only the poles contribute, we show in Supplemental Section II~\cite{Note2} that the operators
\begin{align}
Q_e(z) &=  2\partial_{z^*} \psi_e - v_m \psi_h^\dagger (\cev{\partial}_z-ia)^{m-1} \nonumber\\
Q_h(z) &=  2\partial_{z^*} \psi_h -  v_m  \psi_e^\dagger (\cev{\partial}_z+ia)^{m-1}
\end{align}
satisfy $Q_{e,h}(z)|\Psi_m\rangle=0$.   Here $v_m = 2\pi f/(m-1)!$, and $\cev{\partial}_z$ acts to the left on $\psi^\dagger_{h,e}(z)$ and
\begin{equation}
a(z) = m \int d^2 u \frac{\rho(u)}{i(z-u)}; \quad \rho = \psi^\dagger_e\psi_e - \psi^\dagger_h \psi_h.
\end{equation}
This can be interpreted as $a(z) = a_x - i a_y$, where ${\bf a}$ is a statistical vector potential similar to (\ref{avec}), except with $m$ fluxes per particle, rather than $m-1$.    We then define
\begin{equation}
{\cal H}_m = \frac{1}{2}\int d^2{\bf z} \left[Q_e^\dagger(z) Q_e(z) + Q_h^\dagger(z) Q_h(z)\right].
\label{hexact}
\end{equation}
Since ${\cal H}_m$ is the sum of positive operators, $|\Psi_m\rangle$ is guaranteed to be a ground state. 

For $m=1$,  $Q_{e,h}(z)$ is the Fourier transform of $\sqrt{2E_{\bf k}}\gamma_{{e,h}{\bf k}}$, where $\gamma_{e(h){\bf k}} = u_{\pm\bf k} c_{e(h){\bf k}} \pm v_{\pm\bf k} c^\dagger_{h(e)-{\bf k}}$ are Bogoliubov quasiparticle annihilation operators.  It follows that \eqref{hexact} reduces to \eqref{h2band} and \eqref{hm=1} up to an additive constant.   For $m>1$, \eqref{hexact} involves up to $(2m-1)$ body interactions.   While we have not proven that ${\cal H}_m$ has a gap, it is plausible that it does, provided $|\Psi_m\rangle$ is in the screening phase and has short ranged correlations
~\footnote{In Supplemental Section II\cite{Note2} we also introduce a second set of operators $P_{e,h}(z)$ that annihilate \unexpanded{$|\Psi_m\rangle$} and define a second term in ${\cal H}_m$ that can also contribute to the energy gap.}.
If so, then turning down the several-body interactions will not immediately destroy the state.  This motivates a more practical strategy for realizing this state.

Imagine turning off the interaction terms in \eqref{hexact}, so that $Q_e = 2\partial^*_z\psi_e - v_m \partial^{m-1}_z \psi^\dagger_h$.   This leads to a non-interacting Hamiltonian of the form \eqref{h2band}, where for ${\bf k}\rightarrow 0$
\begin{equation}
\epsilon_{\bf k} = k^2/2; \quad \Delta_{\bf k} = v_m(i k_x + k_y)^m/2^{m-1} . \label{hm}
\end{equation}
This describes a system with quadratically dispersing bands that touch at ${\bf k}=0$ and are coupled by angular momentum $m$ excitonic pairing.    We now argue that this gapless ``$(p+ip)^m$ pairing" state is a candidate for supporting a fractional excitonic insulator in the presence of strong repulsive interactions.

The ground state $|\Phi_m\rangle$ of Eq. \eqref{h2band} with $\epsilon_{\bf k}$ and $\Delta_{\bf k}$ as defined in Eq. \eqref{hm} can be written in the form \eqref{psi g(z-w)}.  Using $g_{\bf k}  \propto (i k_x + k_y)^m/k^2$ for $k\ll \xi^{-1}$ the component with $N$ particles and holes has the form
\begin{equation}
\phi_m^N(\{z_i,w_j\}) = \det\left[g(z_i-w_j)\right]; \quad g(|z|\gg \xi) \propto z^{-m}.
\label{psi0m}
\end{equation}
If we multiply out the determinant and put it over a common denominator, then $\phi_m^N$ gets the denominator in \eqref{wf} right---at least in the universal $z_i-w_j\gg\xi$ limit.   
The numerator of $\phi_m^N$ is not the same as $\psi^N_m$, but if we use the large $z$ limit of $g(z)$ then it will be a degree $m N(N-1)$ polynomial.   As a function of one of its variables (say $z_1$) the numerator has $m(N-1)$ zeros - the same as the numerator of $\psi^N_m$.   $N-1$ of the zeros are guaranteed by Fermi statistics to sit on $z_{2, ..., N}$, but the remaining $(m-1)(N-1)$ zeros are ``wasted" and sit between the particles.    This is similar to a $1/m$ filled Landau level, where the magnetic field guarantees there are $m$ times as many zeros as there are particles.    In that case, repulsive interactions stabilize the Laughlin state, which puts the required zeros on top of the particles.  The above argument strictly applies to the dilute limit, where electrons and holes are separated by more than $\xi$, so $|\Psi_m\rangle$ is in a bound phase.   In the dense limit, however, $|\Psi_m\rangle$ is still more effective than $|\Phi_m\rangle$ at keeping the electrons (holes) apart, and it also builds in the $(p+ip)^m$ pairing of electrons and holes favored by \eqref{hm}.    It will be interesting to test our conjecture that \eqref{hm}, along with strong repulsive interactions can stabilize the fractional excitonic insulator state by the numerical analysis of model systems.

Eq. \eqref{hm} presents an appealing target for band structure engineering.   It requires the crossing of two bands that differ in angular momentum by $m$.   For $m=3$ this can occur at the $\Gamma$ point in a crystal with $C_6$ rotational symmetry but broken time reversal and in-plane mirrors.   For example, this could arise if two bands with $m_j = \pm 3/2$ touch at the Fermi energy.   Here we introduce a simple two band model for spinless electrons that provides a starting point for numerical studies.

Consider a triangular lattice with an $s$ state and a single $f$ state with $m=3$ on each site.  A  Hamiltonian with first and second neighbor hopping can be written as Eq. \eqref{h2band} with
\begin{equation}
\epsilon_{\bf k} = \epsilon_0 - t_0 \gamma_0({\bf k}); \quad \Delta_{\bf k} = t_1 \gamma_1({\bf k}) + i t_2 \gamma_2({\bf k})
\label{htriangle}
\end{equation}
where $\gamma_0({\bf k}) = \sum_n \cos {\bf k}\cdot {\bf a}_{1n}$, $\gamma_1({\bf k}) = \sum_n (-1)^n\sin {\bf k}\cdot  {\bf a}_{1n}$ and $\gamma_2({\bf k}) = \sum_n (-1)^n\sin {\bf k}\cdot  {\bf a}_{2n}$.   Here ${\bf a}_{1(2)n}$ are the $6$ first (second) neighbor lattice vectors at angles $\theta= n\pi /3\ (+ \pi/6)$.   $t_0$ connects nearest neighbors of the same orbitals, while $t_1$ and $t_2$ connect first and second neighbor $s$ and $f$ orbitals with an angle dependent phase $e^{3i\theta}$.   

For $-6 < \epsilon_0/t_0 < 2$ (\ref{htriangle}) is a Chern number $3$ insulator.  Outside that range it is a trivial insulator.    For $\epsilon_0 = 2 t_0$ the gap closes at the 3 $M$ points, while for $\epsilon_0=-6 t_0$ the critical point is at $\Gamma$.  While it is not our primary focus, the Chern number 3 transition is of interest on its own.   For $\epsilon_0 = - 6 t_0 + \delta$ the small $\bf k$ behavior is
\begin{equation}
\epsilon_{\bf k} = \delta + 3 t_0 k^2/2; \quad  \Delta_{\bf k} = t_+ k_+^3 + t_- k_-^3,
\end{equation}
with $t_\pm = (t_1 \pm 3\sqrt{3} t_2)/8$ and $k_\pm = k_x \pm i k_y$.     For $\delta >0$ the gap $E_g \propto \delta$ is at ${\bf k}=0$, but for $\delta<0$ $E_g \propto |\delta|^{3/2}$, and is located on a ``Fermi surface" of radius $\propto |\delta|^{1/2}$.   The critical point $\delta=0$ has precisely the structure of (\ref{hm}) when $t_-=0$~\cite{Note1}.   For non-zero $t_-$, the vorticity 3 winding of $\Delta_{\bf k}$ around ${\bf k}=0$ remains, so the long distance phase winding of $g(z)$ is not altered.   It will be interesting to study this model near the transition to determine whether electron interactions stabilize the fractional excitonic insulator by addressing signatures such as ground state degeneracy, spectral flow under flux insertion and entanglement spectrum.   Importantly, in contrast to the case of fractional Chern insulators, this model should be studied at {\it integer} filling per unit cell.

In this paper we have introduced a paradigm for achieving a FQH state in a correlated fluid of electrons an holes described by a generalization of the Laughlin wavefunction and characterized by $(p_x + i p_y)^m$ excitonic pairing.   This points to several avenues for further investigation.   It will be interesting to numerically study the ground state properties of model  Hamiltonians such as (\ref{htriangle}) with interactions to establish the fractional excitonic insulator phase.  In parallel, it will be interesting to identify materials with band structures that feature an $m=3$ band inversion near the Fermi energy.   Finally, the considerations in this paper can be generalized to describe multi-component systems, superconductors and symmetry protected topological phases.

\acknowledgments

We thank Gene Mele, Ady Stern and Michael Zaletel for helpful discussions.   This work was supported by a Simons Investigator grant from the Simons Foundation and by National Science Foundation Grant DMR-1120901.

\pagebreak
\widetext
\begin{center}
\textbf{\large  Supplemental material for ``Fractional Excitonic Insulator''}

\bigskip
Yichen Hu, J\"orn W. F. Venderbos and C. L. Kane \\
{\it Department of Physics and Astronomy, University of Pennsylvania,
Philadelphia, Pennsylvania 19104}
 \newline
\end{center}
\setcounter{equation}{0}
\setcounter{figure}{0}
\setcounter{table}{0}
\setcounter{page}{1}
\makeatletter
\renewcommand{\theequation}{S\arabic{equation}}
\renewcommand{\thefigure}{S\arabic{figure}}
\renewcommand{\bibnumfmt}[1]{[S#1]}
\renewcommand{\citenumfont}[1]{S#1}

%

\twocolumngrid

\section{Coupled wire construction  \label{sec:wiremodel}}

\begin{figure}
\includegraphics[width=0.9\columnwidth]{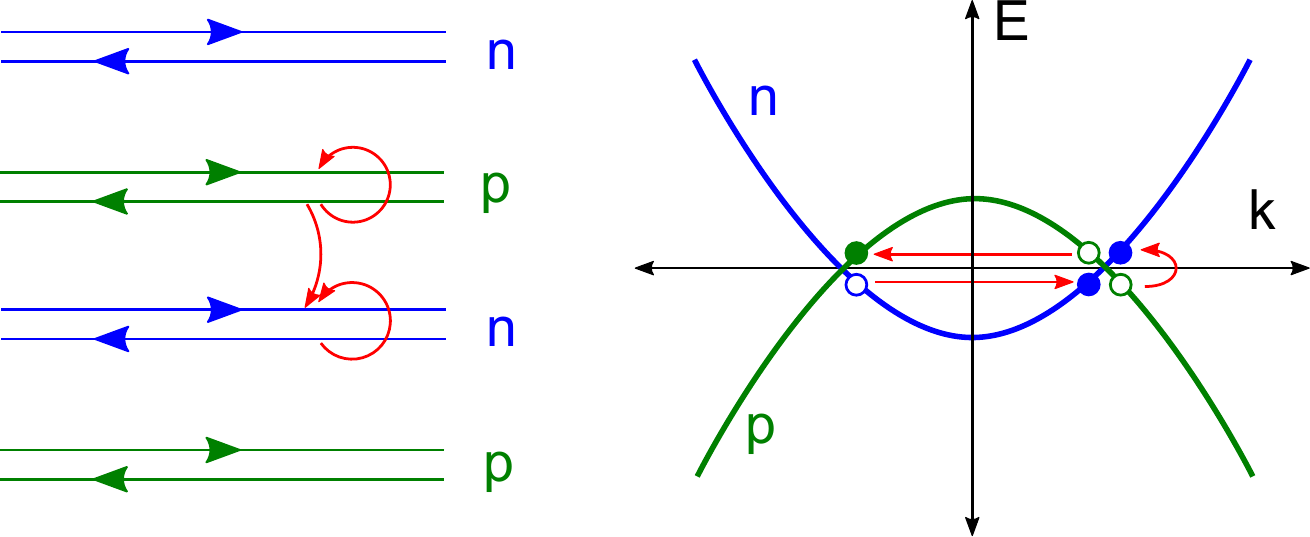}
\caption{(a) An array of alternating $n$-type and $p$-type wires.   (b) Energy bands as a function of momentum, showing the electron like (hole like bands), that live on the odd (even) wires.   The red arrows indicate the correlated tunneling processes that lead to the $\nu = 1/m$ fractional excitonic insulator for the case $m=3$.  }
\label{FigS1}
\end{figure}

In this section we introduce a simple modification of the coupled wire construction \cite{kml2002_SM} that allows us to describe a fractional excitonic insulator at zero magnetic field.   We consider an array of alternating $n$ type and $p$ type wires, as indicated in Fig. \ref{FigS1}.   On the $n$-type wires the right (left) moving states are at momentum $+k_F$ ($-k_F$), but on the $p$-type wires they are at $-k_F$ ($+k_F$).   This allows momentum conserving processes that lead to the quantum Hall effect in zero magnetic field.

Specifically, we consider the Hamiltonian ${\cal H} = {\cal H}_0 + V$, where
\begin{equation}
{\cal H} = -i \sum_i \int dx\psi^\dagger_{i,R} \partial_x \psi_{i,R} - \psi^\dagger_{i,L}\partial_x \psi_{i,L}
\end{equation}
describes the low energy excitations on each wire.   The electron annihilation operator is given by 
\begin{equation}
c_i(x) = e^{\pm i k_F x} \psi_{i,R}  + e^{\mp i k_F x} \psi_{i,L}
\end{equation}
where the upper (lower) sign corresponds to the $n$ type ($p$ type) wires for $i$ odd (even).

The $\nu=1/m$ Laughlin state (for $m$ an odd integer) is generated by introducing the $m$ body coupling term $V^m = \sum_i \int dx (V^m_i(x) + h.c.)$, where 
\begin{equation}
 V_i^m(x) = v_m \psi_{i,R}^{\dagger(m+1)/2} \psi_{i,L}^{(m-1)/2} \psi_{i+1,R}^{\dagger(m-1)/2} \psi_{i+1,L}^{(m+1)/2}
\end{equation}
Here, powers of $\psi_{i,R}$ are understood as an operator product expansion and include appropriate derivatives.   Note that $V^m$ conserves momentum in zero magnetic field for all $m$.   No tuning of the electron or hole densities is required, provided they are equal, so that the Fermi energy is at the band crossing point.   

In the absence of other interactions, $v_m$ has scaling dimension $(1+m^2)/2$, and will be irrelevant for $m>1$.   Nonetheless, as argued in Ref. \onlinecite{kml2002_SM}, it is possible to choose forward scattering interactions that can make any particular $v_m$ relevant.   In the presence of such interactions, $v_m$ will flow to strong coupling, which leads to an energy gap and the $\nu=1/m$ fractional excitonic insulator phase.

The connection with Laughlin's plasma analogy can be understood by considering a particular limit where the problem decouples into independent 1D problems.   When forward scattering interactions on each wire make them Luttinger liquids with $K = 1/m$, the $v_m$ term couples only to a purely chiral operator on each wire.   In this case, $v_m$ is identical to electron tunneling between the edge states of strips of $\nu=1/m$ fractional quantum Hall states, which upon bosonization lead to a $1+1$D sine-Gordon type model.   In this case, $v_m$ has scaling dimension $m$.   Expanding the partition function in powers of $v_m$ leads to exactly the same Coulomb plasma as the analysis of the Laughlin type wavefunction.

\section{Exact Hamiltonian  \label{sec:exactH}}

In this section we demonstrate that the state $\ket{\Psi_m}$ as defined in the main text is the exact ground state wavefunction of Hamiltonian (14) of the main text. Our strategy is to seek operators $X$ which annihilate the ground state, i.e., $X\ket{\Psi_m}=0$. With the help of such operators one may then construct positive (and manifestly Hermitian) operators $\sim X^\dagger X$, which can be used to define a Hamiltonian with $\ket{\Psi_m}$ as its ground state. Any operator $X$ satisfying $X\ket{\Psi_m}=0$ can be used to define a term which may enter in the exact Hamiltonian. In fact, by explicitly constructing two sets of such operators, we will demonstrate that the space of exact Hamiltonians is larger than $\mathcal H_m$ given in the text. The full exact Hamiltonian, which is a sum of all allowed terms, can be used to study more physical few-body pseudopotential Hamiltonians, for which the wavefunction $\ket{\Psi_m}$ may still describe the ground state properties. With this goal in mind, we will conclude this section with a brief comparison to a class exact Hamiltonians for the Laughlin wavefunction introduced in Ref. \onlinecite{kane1991_SM}. In these case of the latter two sets of operators are needed to construct an exact Hamiltonian with the Laughlin state as its ground state and a gap to excited states. 

\subsection{Construction of Hamiltonian  \label{ssec:contruction}}

To begin, first recall that $\ket{\Psi_m}$ is defined as
\be
\ket{\Psi_m} = \sum_N \frac{f^N}{N!} \ket{\Psi^N_m},
\ee
where $\ket{\Psi^N_m}$ is a state with $N$ particle-hole pairs defined as 
\be
\ket{\Psi^N_m} = \int \left( \prod_{i=1}^N dz_i dw_i \right)  \Psi^N_m(\{z_i,w_i\})\ket{N,  \{z_i,w_i\} },
\ee
with $\ket{N,  \{z_i,w_i\} }$ given by $(1/N!)\prod_{i=1}^N\psi_e^\dagger(z_i)\psi_h^\dagger(w_i)\ket{0}$. Note that the factor $1/N!$ ensures proper normalization. 

A natural choice for the annihilation operators involves the derivative operators $\partial_z = \frac12(\partial_z -i \partial_y)$ and $\partial_{z^*} = \frac12(\partial_z +i \partial_y)$. Consider first the derivative operator $\partial_z$. It is a simple matter to verify that the (second quantized) operators 
\beq
P_e(z) &=& (\partial_z - ia)\psi_e(z),  \label{eq:Pe} \\
P_h(z) &=& (\partial_z + ia)\psi_h(z),  \label{eq:Ph}
\eeq
annihilate the states $\ket{\Psi^N_m}$ with $N$ particle-hole pairs, i.e., $P_{e,h}(z) \ket{\Psi^N_m}=0$, where $a=a(z)=a_x-ia_y$ is the statistical gauge field defined as 
\be
ia(z) = m \int d^2u \frac{\rho(u)}{z-u}, \quad \rho = \psi_e^\dagger\psi_e-\psi_h^\dagger\psi_h.  \label{eq:a(z)}
\ee
The statistical gauge field attaches $\pm m$ flux quanta to the particles (holes). Note that within the sector of Fock space defined by $N-1$ electrons and $N$ holes $a(z)$ takes the form
\be
ia(z) =  \sum_{i=1}^{N-1}\frac{m}{z-z_i}-\sum_{i=1}^{N}\frac{m}{z-w_i},
\ee
from which it directly follows that $P_{e,h}(z)$ annihilate each $ \ket{\Psi^N_m}$,  and thus annihilate $\ket{\Psi_m}$. It is worth pointing out that the first quantized operators 
\be
\Pi_\pm \equiv \partial_z \pm ia
\ee 
have the commutators $[\Pi_\pm,\Pi^\dagger_\pm ] = \pm 2\pi \rho = \pm \makebf{\nabla} \times {\bf a}$. We then use the operators $P_{e,h}(z)$ to define the Hamiltonian $\mathcal H^{(1)}_m$ given by
\be
\mathcal H^{(1)}_m = \frac12 \int d^2\bz \left[ P^\dagger_e(z)P_e(z)+P^\dagger_h(z)P_h(z)\right].  \label{eq:H1m}
\ee
By construction this Hamiltonian annihilates the wavefunction, which implies that $\ket{\Psi_m}$ is eigenstate with eigenvalue $0$. Since $\mathcal H^{(1)}_m$ is positive $\ket{\Psi_m}$ must be a ground state.

Next, consider the derivative operator $\partial_{z^*}$. Since only the holomorphic coordinates $z_i$ enter the wavefunction, the action of $\partial_{z^*}$ requires a more careful treatment. We first define and evaluate
\begin{equation} 
\Phi(z)=\partial_{z^*}  \psi_e(z) \ket{\Psi^N_m}.   \label{eq:Phi}
\end{equation}
Note that $\psi_e(z)$ picks one of the $N$ electron coordinates $z_i$ and sets it to $z$. Furthermore, $\Phi(z)$ describes a state with one electron removed from $ \ket{\Psi_N}$, which we may alternatively view as a state with one hole added to $ \ket{\Psi^{N-1}_m}$. We will therefore seek to relate $\Phi(z)$ to $\ket{\Psi^{N-1}_m}$.

Note that $\partial_{z^*}$ gives zero when acting on an analytic function except at the poles. Since there are $N$ such poles, located at $w_i$, we can rename each pole $w$ and then relabel the remaining $N-1$ $w_i$'s. We separate out the dependence on $z$ and $w$ and obtain
\be
\psi_e(z) \ket{\Psi^N_m}=N \int dw \frac{1}{(z-w)^m} \psi_h^{\dagger}(w) \ket{\widetilde \Psi^{N-1}_m(z,w)},
\ee
where the state $ \ket{\widetilde \Psi^{N-1}_m(z,w)}$ is defined as 
\begin{multline}
 \ket{\widetilde \Psi^{N-1}_m(z,w)} = \int  \left( \prod_{i=1}^{N-1} dz_i dw_i \right) \widetilde \Psi^{N-1}_m(z,w,\{z_i,w_i\}) \\
\times  \ket{N-1,  \{z_i,w_i\} } \label{eq:tilde psi}
\end{multline}
with a wavefunction given by
\be
\widetilde \Psi^{N-1}_m(z,w,\{z_i,w_i\})=  F(z,w,\{z_i,w_i\}) \Psi^{N-1}_m(\{z_i,w_i\}).
\ee
Here $\Psi^{N-1}_m(\{z_i,w_i\})$ is the wave function of $ \ket{\Psi^{N-1}_m}$ and $F$ defined as
\begin{equation}
F(z,w,\{z_i,w_i\})=\prod_i^{N-1} \frac{(z-z_i)^m(w-w_i)^m}{(w-z_i)^m(z-w_i)^m}.
\end{equation}
We observe that $F$ can be rewritten as 
\beq
F &=& e^{m\sum \log{\frac{z-z_i}{w-z_i}}-\log{\frac{z-w_i}{w-w_i}}} \nonumber \\
   &=& e^{\int_w^z du \sum_i(\frac{m}{u-z_i}-\frac{m}{u-w_i})} =e^{i \int_w^z du a(u)},
\eeq 
where $a(u)$ is the gauge field introduced in \eqref{eq:a(z)} and is given by
\be
ia(u) =   \sum_{i=1}^{N-1} \left(\frac{m}{u-z_i}-\frac{m}{u-w_i}\right).
\ee
We thus find that the Eq. \eqref{eq:tilde psi} can be expressed in the concise form
\be
 \ket{\widetilde \Psi^{N-1}_m(z,w)} =  e^{\int_w^z du a(u)} \ket{ \Psi^{N-1}_m}. 
\ee

The next step is to consider the action of $\partial_{z^*}$ on the pole at $w$. One finds that
\beq
\partial_{z^*} \frac{1}{(z-w)^m}  &= & \frac{\partial_w^{m-1}}{(m-1)!}\partial_{z^*} \frac{1}{z-w} \nonumber \\
&=&\frac{\pi}{(m-1)!}\partial_w^{m-1} \delta^{(2)} (z-w)
\eeq
where we used Cauchy's integral formula and 
\be
\partial_{z^*} \frac{1}{z-w}=\pi \delta^{(2)} (z-w)
\ee
As a result, $\Phi(z)$ defined in Eq. \eqref{eq:Phi} becomes 
\begin{multline}
\Phi(z)=\frac{\pi N}{(m-1)!}\int dw [\partial^{m-1}_w \delta(z-w)] \psi_h^{\dagger}(w) \\
\times   e^{i \int_w^z du a(u)}\ket{\Psi_{N-1}}
\end{multline}
The right hand side can be integrated by parts to obtain
\be
\partial_{z^*} \psi_e(z) \ket{\Psi_N}=\frac{ \pi N}{(m-1)!}\psi_h^{\dagger}(\overleftarrow{\partial_z} -ia)^{m-1}  \ket{\Psi_{N-1}}.  \label{eq:relation}
\ee
Equation \eqref{eq:relation} gives the desired relation between $\Phi(z)$ and $\ket{\Psi_{N-1}}$, and we use it to define the operator
\be
Q_e(z)=  \partial_{z^*} \psi_e(z) - \frac{ \pi f}{(m-1)!}\psi_h^{\dagger}(\overleftarrow{\partial_z} -ia)^{m-1},  \label{eq:Qe}
\ee
which, by construction $Q_e(z)$, annihilates the state $\ket{\Psi_m}$. A very similar analysis can be applied to $ \partial_{z^*}\psi_h(z)$ and leads to the definition of $Q_h(z)$, which is given by \eqref{eq:Qe} after exchanging $\psi_e,\psi_e^{\dagger} \leftrightarrow \psi_h, \psi_h^{\dagger}$ and substituting $a \to -a$. 

We use the operators $Q_{e,h}(z)$ to construct another positive Hermitian $\mathcal H^{(2)}_m$ given by
\be
\mathcal H^{(2)}_m = \frac12 \int d^2\bz \left[ Q^\dagger_e(z)Q_e(z)+Q^\dagger_h(z)Q_h(z)\right],  \label{eq:H2m}
\ee
which has $\ket{\Psi_m}$ as a zero energy ground state. Combining Eqs. \eqref{eq:H1m} and \eqref{eq:H2m}, one may form the exact Hamiltonian
\be
\mathcal H_m = \lambda_1 \mathcal H^{(1)}_m+\lambda_2 \mathcal H^{(2)}_m.  \label{eq:Hm}
\ee
Note that in the case $m=1$ the $\mathcal H^{(2)}_{m=1}$ simply reduces to the non-interacting for $p_x+ip_x$ excitonic pairing, see Eqs. (3) and (4) of the main text. This implies that for $m=1$ the exact Hamiltonian $\mathcal H_m$ is specified by $(\lambda_1,\lambda_2)=(0,1)$. 



\subsection{Comparison to lowest Landau level  \label{ssec:landaulevel}}

Let us now compare the operators $Q_{e,h}$ and $P_{e,h}$ to operators which annihilate the Laughlin wavefunction $\Psi^{\text{Laughlin}}_m $ describing a fractional quantum Hall liquid in the lowest Landau level at filling factor $\nu=1/m$. In the symmetric gauge the single-particle states in the lowest Landau level are eigenstates of angular momentum. 
The Laughlin wave function takes the form
\be
\Psi^{\text{Laughlin}}_m \propto \prod_{i<j}(z_i-z_j)^m e^{- \sum_i z_i z_i^*}.  \label{eq:Laughlin}
\ee
It is worth pointing out that the Gaussian piece originates from the magnetic field (and we have taken twice the magnetic length as the unit of length). 

Now consider the following two (first-quantized) operators involving the derivatives $\partial_{z}$ and $\partial_{z^*}$:
\beq
\Pi &=& \partial_{z^*} + z ,  \\
\Lambda   &=& \partial_z + z^* - ia .
\eeq 
The operator $\Pi $ annihilates all single-particle states of the lowest Landau level and therefore annihilates $\Psi^{\text{Laughlin}}_m$. This can be understood by recognizing that $\Pi$ and $\Pi^\dagger$ are the ladder operators of the Landau levels, i.e., $\Pi$ ($\Pi^\dagger$) lowers (raises) the Landau level index. The Hamiltonian constructed from $\Pi $, given by $\Pi^\dagger \Pi$, simply corresponds to the kinetic energy of a particle in a magnetic field (up to an additive constant). Therefore, $\Pi^\dagger \Pi$ does not by itself lead to energy gap at filling $\nu=1/m$. One may also note that since $\Pi^\dagger \Pi$ annihilates all wavefunctions constructed from states in the lowest Landau level, it is certainly not sufficient to single out the Laughlin wavefunction as the ground state wavefunction.  

Instead, the Laughlin wavefunction is selected by interactions, and an exact interacting Hamiltonian can be constructed by including $\Lambda^\dagger\Lambda$. Note that $\partial_z + z^*$ lowers the angular momentum of the single-particle states, i.e., $\Lambda$ is defined as the angular momentum lowering operator minus a statistical gauge field $a$ given by
\be
 ia(z_i) =  m\sum_{j\neq i} \frac{1}{z_i-z_j}.
\ee
It is then straightforward to verify that $\Lambda$ indeed annihilates the Laughlin state. As a result, in the fractional quantum Hall problem both operators $\Pi $ and $\Lambda $ are needed to construct an exact interacting Hamilonian with \eqref{eq:Laughlin} as its ground state wave function. Such Hamiltonian can be related to an interacting Hamiltonian with short-ranged two-body interactions \cite{trugman1985_SM}, of which the ground state properties are described by \eqref{eq:Laughlin}.

This leads to the expectation that in the case of the fractional excitonic insulator a general Hamiltonian of the form \eqref{eq:Hm}, involving both the $Q_{e,h}$ and $P_{e,h}$ operators, should be considered for $m\neq 1$.


\begin{thebibliography}{40}%
\makeatletter
\providecommand \@ifxundefined [1]{%
 \@ifx{#1\undefined}
}%
\providecommand \@ifnum [1]{%
 \ifnum #1\expandafter \@firstoftwo
 \else \expandafter \@secondoftwo
 \fi
}%
\providecommand \@ifx [1]{%
 \ifx #1\expandafter \@firstoftwo
 \else \expandafter \@secondoftwo
 \fi
}%
\providecommand \natexlab [1]{#1}%
\providecommand \enquote  [1]{``#1''}%
\providecommand \bibnamefont  [1]{#1}%
\providecommand \bibfnamefont [1]{#1}%
\providecommand \citenamefont [1]{#1}%
\providecommand \href@noop [0]{\@secondoftwo}%
\providecommand \href [0]{\begingroup \@sanitize@url \@href}%
\providecommand \@href[1]{\@@startlink{#1}\@@href}%
\providecommand \@@href[1]{\endgroup#1\@@endlink}%
\providecommand \@sanitize@url [0]{\catcode `\\12\catcode `\$12\catcode
  `\&12\catcode `\#12\catcode `\^12\catcode `\_12\catcode `\%12\relax}%
\providecommand \@@startlink[1]{}%
\providecommand \@@endlink[0]{}%
\providecommand \url  [0]{\begingroup\@sanitize@url \@url }%
\providecommand \@url [1]{\endgroup\@href {#1}{\urlprefix }}%
\providecommand \urlprefix  [0]{URL }%
\providecommand \Eprint [0]{\href }%
\providecommand \doibase [0]{http://dx.doi.org/}%
\providecommand \selectlanguage [0]{\@gobble}%
\providecommand \bibinfo  [0]{\@secondoftwo}%
\providecommand \bibfield  [0]{\@secondoftwo}%
\providecommand \translation [1]{[#1]}%
\providecommand \BibitemOpen [0]{}%
\providecommand \bibitemStop [0]{}%
\providecommand \bibitemNoStop [0]{.\EOS\space}%
\providecommand \EOS [0]{\spacefactor3000\relax}%
\providecommand \BibitemShut  [1]{\csname bibitem#1\endcsname}%
\let\auto@bib@innerbib\@empty
\bibitem [{\citenamefont {Klitzing}\ \emph {et~al.}(1980)\citenamefont
  {Klitzing}, \citenamefont {Dorda},\ and\ \citenamefont
  {Pepper}}]{klitzing1980}%
  \BibitemOpen
  \bibfield  {author} {\bibinfo {author} {\bibfnamefont {K.~v.}\ \bibnamefont
  {Klitzing}}, \bibinfo {author} {\bibfnamefont {G.}~\bibnamefont {Dorda}}, \
  and\ \bibinfo {author} {\bibfnamefont {M.}~\bibnamefont {Pepper}},\ }\href
  {\doibase 10.1103/PhysRevLett.45.494} {\bibfield  {journal} {\bibinfo
  {journal} {Phys. Rev. Lett.}\ }\textbf {\bibinfo {volume} {45}},\ \bibinfo
  {pages} {494} (\bibinfo {year} {1980})}\BibitemShut {NoStop}%
\bibitem [{\citenamefont {Thouless}\ \emph {et~al.}(1982)\citenamefont
  {Thouless}, \citenamefont {Kohmoto}, \citenamefont {Nightingale},\ and\
  \citenamefont {den Nijs}}]{tknn1982}%
  \BibitemOpen
  \bibfield  {author} {\bibinfo {author} {\bibfnamefont {D.~J.}\ \bibnamefont
  {Thouless}}, \bibinfo {author} {\bibfnamefont {M.}~\bibnamefont {Kohmoto}},
  \bibinfo {author} {\bibfnamefont {M.~P.}\ \bibnamefont {Nightingale}}, \ and\
  \bibinfo {author} {\bibfnamefont {M.}~\bibnamefont {den Nijs}},\ }\href
  {\doibase 10.1103/PhysRevLett.49.405} {\bibfield  {journal} {\bibinfo
  {journal} {Phys. Rev. Lett.}\ }\textbf {\bibinfo {volume} {49}},\ \bibinfo
  {pages} {405} (\bibinfo {year} {1982})}\BibitemShut {NoStop}%
\bibitem [{\citenamefont {Hofstadter}(1976)}]{hofstadter1976}%
  \BibitemOpen
  \bibfield  {author} {\bibinfo {author} {\bibfnamefont {D.~R.}\ \bibnamefont
  {Hofstadter}},\ }\href {\doibase 10.1103/PhysRevB.14.2239} {\bibfield
  {journal} {\bibinfo  {journal} {Phys. Rev. B}\ }\textbf {\bibinfo {volume}
  {14}},\ \bibinfo {pages} {2239} (\bibinfo {year} {1976})}\BibitemShut
  {NoStop}%
\bibitem [{\citenamefont {Haldane}(1988)}]{haldane1988}%
  \BibitemOpen
  \bibfield  {author} {\bibinfo {author} {\bibfnamefont {F.~D.~M.}\
  \bibnamefont {Haldane}},\ }\href {\doibase 10.1103/PhysRevLett.61.2015}
  {\bibfield  {journal} {\bibinfo  {journal} {Phys. Rev. Lett.}\ }\textbf
  {\bibinfo {volume} {61}},\ \bibinfo {pages} {2015} (\bibinfo {year}
  {1988})}\BibitemShut {NoStop}%
\bibitem [{\citenamefont {Ludwig}\ \emph {et~al.}(1994)\citenamefont {Ludwig},
  \citenamefont {Fisher}, \citenamefont {Shankar},\ and\ \citenamefont
  {Grinstein}}]{ludwig1994}%
  \BibitemOpen
  \bibfield  {author} {\bibinfo {author} {\bibfnamefont {A.~W.~W.}\
  \bibnamefont {Ludwig}}, \bibinfo {author} {\bibfnamefont {M.~P.~A.}\
  \bibnamefont {Fisher}}, \bibinfo {author} {\bibfnamefont {R.}~\bibnamefont
  {Shankar}}, \ and\ \bibinfo {author} {\bibfnamefont {G.}~\bibnamefont
  {Grinstein}},\ }\href {\doibase 10.1103/PhysRevB.50.7526} {\bibfield
  {journal} {\bibinfo  {journal} {Phys. Rev. B}\ }\textbf {\bibinfo {volume}
  {50}},\ \bibinfo {pages} {7526} (\bibinfo {year} {1994})}\BibitemShut
  {NoStop}%
\bibitem [{\citenamefont {Read}\ and\ \citenamefont
  {Green}(2000)}]{readgreen2000}%
  \BibitemOpen
  \bibfield  {author} {\bibinfo {author} {\bibfnamefont {N.}~\bibnamefont
  {Read}}\ and\ \bibinfo {author} {\bibfnamefont {D.}~\bibnamefont {Green}},\
  }\href {\doibase 10.1103/PhysRevB.61.10267} {\bibfield  {journal} {\bibinfo
  {journal} {Phys. Rev. B}\ }\textbf {\bibinfo {volume} {61}},\ \bibinfo
  {pages} {10267} (\bibinfo {year} {2000})}\BibitemShut {NoStop}%
\bibitem [{\citenamefont {Bernevig}\ \emph {et~al.}(2006)\citenamefont
  {Bernevig}, \citenamefont {Hughes},\ and\ \citenamefont {Zhang}}]{bhz2006}%
  \BibitemOpen
  \bibfield  {author} {\bibinfo {author} {\bibfnamefont {B.~A.}\ \bibnamefont
  {Bernevig}}, \bibinfo {author} {\bibfnamefont {T.~L.}\ \bibnamefont
  {Hughes}}, \ and\ \bibinfo {author} {\bibfnamefont {S.-C.}\ \bibnamefont
  {Zhang}},\ }\href {\doibase 10.1126/science.1133734} {\bibfield  {journal}
  {\bibinfo  {journal} {Science}\ }\textbf {\bibinfo {volume} {314}},\ \bibinfo
  {pages} {1757} (\bibinfo {year} {2006})}\BibitemShut {NoStop}%
\bibitem [{\citenamefont {Fu}\ and\ \citenamefont {Kane}(2007)}]{fukane2007}%
  \BibitemOpen
  \bibfield  {author} {\bibinfo {author} {\bibfnamefont {L.}~\bibnamefont
  {Fu}}\ and\ \bibinfo {author} {\bibfnamefont {C.~L.}\ \bibnamefont {Kane}},\
  }\href {\doibase 10.1103/PhysRevB.76.045302} {\bibfield  {journal} {\bibinfo
  {journal} {Phys. Rev. B}\ }\textbf {\bibinfo {volume} {76}},\ \bibinfo
  {pages} {045302} (\bibinfo {year} {2007})}\BibitemShut {NoStop}%
\bibitem [{\citenamefont {Yu}\ \emph {et~al.}(2010)\citenamefont {Yu},
  \citenamefont {Zhang}, \citenamefont {Zhang}, \citenamefont {Zhang},
  \citenamefont {Dai},\ and\ \citenamefont {Fang}}]{yu2010}%
  \BibitemOpen
  \bibfield  {author} {\bibinfo {author} {\bibfnamefont {R.}~\bibnamefont
  {Yu}}, \bibinfo {author} {\bibfnamefont {W.}~\bibnamefont {Zhang}}, \bibinfo
  {author} {\bibfnamefont {H.-J.}\ \bibnamefont {Zhang}}, \bibinfo {author}
  {\bibfnamefont {S.-C.}\ \bibnamefont {Zhang}}, \bibinfo {author}
  {\bibfnamefont {X.}~\bibnamefont {Dai}}, \ and\ \bibinfo {author}
  {\bibfnamefont {Z.}~\bibnamefont {Fang}},\ }\href {\doibase
  10.1126/science.1187485} {\bibfield  {journal} {\bibinfo  {journal}
  {Science}\ }\textbf {\bibinfo {volume} {329}},\ \bibinfo {pages} {61}
  (\bibinfo {year} {2010})}\BibitemShut {NoStop}%
\bibitem [{\citenamefont {Tang}\ \emph {et~al.}(2011)\citenamefont {Tang},
  \citenamefont {Mei},\ and\ \citenamefont {Wen}}]{tang2011}%
  \BibitemOpen
  \bibfield  {author} {\bibinfo {author} {\bibfnamefont {E.}~\bibnamefont
  {Tang}}, \bibinfo {author} {\bibfnamefont {J.-W.}\ \bibnamefont {Mei}}, \
  and\ \bibinfo {author} {\bibfnamefont {X.-G.}\ \bibnamefont {Wen}},\ }\href
  {\doibase 10.1103/PhysRevLett.106.236802} {\bibfield  {journal} {\bibinfo
  {journal} {Phys. Rev. Lett.}\ }\textbf {\bibinfo {volume} {106}},\ \bibinfo
  {pages} {236802} (\bibinfo {year} {2011})}\BibitemShut {NoStop}%
\bibitem [{\citenamefont {Neupert}\ \emph {et~al.}(2011)\citenamefont
  {Neupert}, \citenamefont {Santos}, \citenamefont {Ryu}, \citenamefont
  {Chamon},\ and\ \citenamefont {Mudry}}]{neupert2011}%
  \BibitemOpen
  \bibfield  {author} {\bibinfo {author} {\bibfnamefont {T.}~\bibnamefont
  {Neupert}}, \bibinfo {author} {\bibfnamefont {L.}~\bibnamefont {Santos}},
  \bibinfo {author} {\bibfnamefont {S.}~\bibnamefont {Ryu}}, \bibinfo {author}
  {\bibfnamefont {C.}~\bibnamefont {Chamon}}, \ and\ \bibinfo {author}
  {\bibfnamefont {C.}~\bibnamefont {Mudry}},\ }\href {\doibase
  10.1103/PhysRevB.84.165107} {\bibfield  {journal} {\bibinfo  {journal} {Phys.
  Rev. B}\ }\textbf {\bibinfo {volume} {84}},\ \bibinfo {pages} {165107}
  (\bibinfo {year} {2011})}\BibitemShut {NoStop}%
\bibitem [{\citenamefont {Sun}\ \emph {et~al.}(2011)\citenamefont {Sun},
  \citenamefont {Gu}, \citenamefont {Katsura},\ and\ \citenamefont
  {Das~Sarma}}]{sun2011}%
  \BibitemOpen
  \bibfield  {author} {\bibinfo {author} {\bibfnamefont {K.}~\bibnamefont
  {Sun}}, \bibinfo {author} {\bibfnamefont {Z.}~\bibnamefont {Gu}}, \bibinfo
  {author} {\bibfnamefont {H.}~\bibnamefont {Katsura}}, \ and\ \bibinfo
  {author} {\bibfnamefont {S.}~\bibnamefont {Das~Sarma}},\ }\href {\doibase
  10.1103/PhysRevLett.106.236803} {\bibfield  {journal} {\bibinfo  {journal}
  {Phys. Rev. Lett.}\ }\textbf {\bibinfo {volume} {106}},\ \bibinfo {pages}
  {236803} (\bibinfo {year} {2011})}\BibitemShut {NoStop}%
\bibitem [{\citenamefont {Regnault}\ and\ \citenamefont
  {Bernevig}(2011)}]{regnault2011}%
  \BibitemOpen
  \bibfield  {author} {\bibinfo {author} {\bibfnamefont {N.}~\bibnamefont
  {Regnault}}\ and\ \bibinfo {author} {\bibfnamefont {B.~A.}\ \bibnamefont
  {Bernevig}},\ }\href {\doibase 10.1103/PhysRevX.1.021014} {\bibfield
  {journal} {\bibinfo  {journal} {Phys. Rev. X}\ }\textbf {\bibinfo {volume}
  {1}},\ \bibinfo {pages} {021014} (\bibinfo {year} {2011})}\BibitemShut
  {NoStop}%
\bibitem [{\citenamefont {Parameswaran}\ \emph {et~al.}(2013)\citenamefont
  {Parameswaran}, \citenamefont {Roy},\ and\ \citenamefont
  {Sondhi}}]{parameswaran2013}%
  \BibitemOpen
  \bibfield  {author} {\bibinfo {author} {\bibfnamefont {S.~A.}\ \bibnamefont
  {Parameswaran}}, \bibinfo {author} {\bibfnamefont {R.}~\bibnamefont {Roy}}, \
  and\ \bibinfo {author} {\bibfnamefont {S.~L.}\ \bibnamefont {Sondhi}},\
  }\href {\doibase 10.1016/j.crhy.2013.04.003} {\bibfield  {journal} {\bibinfo
  {journal} {Comptes Rendus Physique}\ }\textbf {\bibinfo {volume} {14}},\
  \bibinfo {pages} {816} (\bibinfo {year} {2013})}\BibitemShut {NoStop}%
\bibitem [{\citenamefont {Neupert}\ \emph {et~al.}(2015)\citenamefont
  {Neupert}, \citenamefont {Chamon}, \citenamefont {Iadecola}, \citenamefont
  {Santos},\ and\ \citenamefont {Mudry}}]{neupert2015}%
  \BibitemOpen
  \bibfield  {author} {\bibinfo {author} {\bibfnamefont {T.}~\bibnamefont
  {Neupert}}, \bibinfo {author} {\bibfnamefont {C.}~\bibnamefont {Chamon}},
  \bibinfo {author} {\bibfnamefont {T.}~\bibnamefont {Iadecola}}, \bibinfo
  {author} {\bibfnamefont {L.~H.}\ \bibnamefont {Santos}}, \ and\ \bibinfo
  {author} {\bibfnamefont {C.}~\bibnamefont {Mudry}},\ }\href {\doibase
  10.1088/0031-8949/2015/T164/014005} {\bibfield  {journal} {\bibinfo
  {journal} {Physica Scripta}\ }\textbf {\bibinfo {volume} {2015}},\ \bibinfo
  {pages} {014005} (\bibinfo {year} {2015})}\BibitemShut {NoStop}%
\bibitem [{\citenamefont {Liu}\ and\ \citenamefont
  {Bergholtz}(2013)}]{liu2013review}%
  \BibitemOpen
  \bibfield  {author} {\bibinfo {author} {\bibfnamefont {Z.}~\bibnamefont
  {Liu}}\ and\ \bibinfo {author} {\bibfnamefont {E.~J.}\ \bibnamefont
  {Bergholtz}},\ }\href {\doibase 10.1142/S021797921330017X} {\bibfield
  {journal} {\bibinfo  {journal} {Int. J. Mod. Phys. B}\ }\textbf {\bibinfo
  {volume} {27}},\ \bibinfo {pages} {1330017} (\bibinfo {year}
  {2013})}\BibitemShut {NoStop}%
\bibitem [{\citenamefont {Spanton}\ \emph {et~al.}(2018)\citenamefont
  {Spanton}, \citenamefont {Zibrov}, \citenamefont {Zhou}, \citenamefont
  {Taniguchi}, \citenamefont {Watanabe}, \citenamefont {Zaletel},\ and\
  \citenamefont {Young}}]{spanton2018}%
  \BibitemOpen
  \bibfield  {author} {\bibinfo {author} {\bibfnamefont {E.~M.}\ \bibnamefont
  {Spanton}}, \bibinfo {author} {\bibfnamefont {A.~A.}\ \bibnamefont {Zibrov}},
  \bibinfo {author} {\bibfnamefont {H.}~\bibnamefont {Zhou}}, \bibinfo {author}
  {\bibfnamefont {T.}~\bibnamefont {Taniguchi}}, \bibinfo {author}
  {\bibfnamefont {K.}~\bibnamefont {Watanabe}}, \bibinfo {author}
  {\bibfnamefont {M.~P.}\ \bibnamefont {Zaletel}}, \ and\ \bibinfo {author}
  {\bibfnamefont {A.~F.}\ \bibnamefont {Young}},\ }\href {\doibase
  10.1126/science.aan8458} {\bibfield  {journal} {\bibinfo  {journal}
  {Science}\ }\textbf {\bibinfo {volume} {360}},\ \bibinfo {pages} {62}
  (\bibinfo {year} {2018})}\BibitemShut {NoStop}%
\bibitem [{\citenamefont {Laughlin}(1983)}]{laughlin1983}%
  \BibitemOpen
  \bibfield  {author} {\bibinfo {author} {\bibfnamefont {R.~B.}\ \bibnamefont
  {Laughlin}},\ }\href {\doibase 10.1103/PhysRevLett.50.1395} {\bibfield
  {journal} {\bibinfo  {journal} {Phys. Rev. Lett.}\ }\textbf {\bibinfo
  {volume} {50}},\ \bibinfo {pages} {1395} (\bibinfo {year}
  {1983})}\BibitemShut {NoStop}%
\bibitem [{\citenamefont {Halperin}(1983)}]{halperin1983}%
  \BibitemOpen
  \bibfield  {author} {\bibinfo {author} {\bibfnamefont {B.~I.}\ \bibnamefont
  {Halperin}},\ }\href@noop {} {\bibfield  {journal} {\bibinfo  {journal}
  {Helv. Phys. Acta}\ }\textbf {\bibinfo {volume} {56}},\ \bibinfo {pages} {75}
  (\bibinfo {year} {1983})}\BibitemShut {NoStop}%
\bibitem [{\citenamefont {Kim}\ \emph {et~al.}(2001)\citenamefont {Kim},
  \citenamefont {Nayak}, \citenamefont {Demler}, \citenamefont {Read},\ and\
  \citenamefont {Das~Sarma}}]{kim2001}%
  \BibitemOpen
  \bibfield  {author} {\bibinfo {author} {\bibfnamefont {Y.~B.}\ \bibnamefont
  {Kim}}, \bibinfo {author} {\bibfnamefont {C.}~\bibnamefont {Nayak}}, \bibinfo
  {author} {\bibfnamefont {E.}~\bibnamefont {Demler}}, \bibinfo {author}
  {\bibfnamefont {N.}~\bibnamefont {Read}}, \ and\ \bibinfo {author}
  {\bibfnamefont {S.}~\bibnamefont {Das~Sarma}},\ }\href {\doibase
  10.1103/PhysRevB.63.205315} {\bibfield  {journal} {\bibinfo  {journal} {Phys.
  Rev. B}\ }\textbf {\bibinfo {volume} {63}},\ \bibinfo {pages} {205315}
  (\bibinfo {year} {2001})}\BibitemShut {NoStop}%
\bibitem [{Note1()}]{Note1}%
  \BibitemOpen
  \bibinfo {note} {The phase of $\Delta _{\protect \bf k}$ can be chosen by
  defining the phase of $c_{h{\protect \bf k}}$. The choice in (\ref {hm=1})
  makes $f$ in (\ref {psin}) real.}\BibitemShut {Stop}%
\bibitem [{\citenamefont {Liu}\ \emph {et~al.}(2010)\citenamefont {Liu},
  \citenamefont {Liu}, \citenamefont {Wu},\ and\ \citenamefont
  {Sinova}}]{liu2010}%
  \BibitemOpen
  \bibfield  {author} {\bibinfo {author} {\bibfnamefont {X.-J.}\ \bibnamefont
  {Liu}}, \bibinfo {author} {\bibfnamefont {X.}~\bibnamefont {Liu}}, \bibinfo
  {author} {\bibfnamefont {C.}~\bibnamefont {Wu}}, \ and\ \bibinfo {author}
  {\bibfnamefont {J.}~\bibnamefont {Sinova}},\ }\href {\doibase
  10.1103/PhysRevA.81.033622} {\bibfield  {journal} {\bibinfo  {journal} {Phys.
  Rev. A}\ }\textbf {\bibinfo {volume} {81}},\ \bibinfo {pages} {033622}
  (\bibinfo {year} {2010})}\BibitemShut {NoStop}%
\bibitem [{\citenamefont {Liu}\ \emph {et~al.}(2011)\citenamefont {Liu},
  \citenamefont {Wang}, \citenamefont {Xie},\ and\ \citenamefont
  {Yu}}]{liu2011}%
  \BibitemOpen
  \bibfield  {author} {\bibinfo {author} {\bibfnamefont {X.}~\bibnamefont
  {Liu}}, \bibinfo {author} {\bibfnamefont {Z.}~\bibnamefont {Wang}}, \bibinfo
  {author} {\bibfnamefont {X.~C.}\ \bibnamefont {Xie}}, \ and\ \bibinfo
  {author} {\bibfnamefont {Y.}~\bibnamefont {Yu}},\ }\href {\doibase
  10.1103/PhysRevB.83.125105} {\bibfield  {journal} {\bibinfo  {journal} {Phys.
  Rev. B}\ }\textbf {\bibinfo {volume} {83}},\ \bibinfo {pages} {125105}
  (\bibinfo {year} {2011})}\BibitemShut {NoStop}%
\bibitem [{\citenamefont {Milovanovi\ifmmode~\acute{c}\else \'{c}\fi{}}\ and\
  \citenamefont {Read}(1996)}]{milovanovic1995}%
  \BibitemOpen
  \bibfield  {author} {\bibinfo {author} {\bibfnamefont {M.}~\bibnamefont
  {Milovanovi\ifmmode~\acute{c}\else \'{c}\fi{}}}\ and\ \bibinfo {author}
  {\bibfnamefont {N.}~\bibnamefont {Read}},\ }\href {\doibase
  10.1103/PhysRevB.53.13559} {\bibfield  {journal} {\bibinfo  {journal} {Phys.
  Rev. B}\ }\textbf {\bibinfo {volume} {53}},\ \bibinfo {pages} {13559}
  (\bibinfo {year} {1996})}\BibitemShut {NoStop}%
\bibitem [{\citenamefont {Yang}(2001)}]{yang2001}%
  \BibitemOpen
  \bibfield  {author} {\bibinfo {author} {\bibfnamefont {K.}~\bibnamefont
  {Yang}},\ }\href {\doibase 10.1103/PhysRevLett.87.056802} {\bibfield
  {journal} {\bibinfo  {journal} {Phys. Rev. Lett.}\ }\textbf {\bibinfo
  {volume} {87}},\ \bibinfo {pages} {056802} (\bibinfo {year}
  {2001})}\BibitemShut {NoStop}%
\bibitem [{\citenamefont {J\'erome}\ \emph {et~al.}(1967)\citenamefont
  {J\'erome}, \citenamefont {Rice},\ and\ \citenamefont {Kohn}}]{jerome1967}%
  \BibitemOpen
  \bibfield  {author} {\bibinfo {author} {\bibfnamefont {D.}~\bibnamefont
  {J\'erome}}, \bibinfo {author} {\bibfnamefont {T.~M.}\ \bibnamefont {Rice}},
  \ and\ \bibinfo {author} {\bibfnamefont {W.}~\bibnamefont {Kohn}},\ }\href
  {\doibase 10.1103/PhysRev.158.462} {\bibfield  {journal} {\bibinfo  {journal}
  {Phys. Rev.}\ }\textbf {\bibinfo {volume} {158}},\ \bibinfo {pages} {462}
  (\bibinfo {year} {1967})}\BibitemShut {NoStop}%
\bibitem [{\citenamefont {Halperin}\ and\ \citenamefont
  {Rice}(1968)}]{halperin1968}%
  \BibitemOpen
  \bibfield  {author} {\bibinfo {author} {\bibfnamefont {B.~I.}\ \bibnamefont
  {Halperin}}\ and\ \bibinfo {author} {\bibfnamefont {T.~M.}\ \bibnamefont
  {Rice}},\ }\href {\doibase 10.1103/RevModPhys.40.755} {\bibfield  {journal}
  {\bibinfo  {journal} {Rev. Mod. Phys.}\ }\textbf {\bibinfo {volume} {40}},\
  \bibinfo {pages} {755} (\bibinfo {year} {1968})}\BibitemShut {NoStop}%
\bibitem [{\citenamefont {Halperin}\ \emph {et~al.}(1993)\citenamefont
  {Halperin}, \citenamefont {Lee},\ and\ \citenamefont {Read}}]{hlr1993}%
  \BibitemOpen
  \bibfield  {author} {\bibinfo {author} {\bibfnamefont {B.~I.}\ \bibnamefont
  {Halperin}}, \bibinfo {author} {\bibfnamefont {P.~A.}\ \bibnamefont {Lee}}, \
  and\ \bibinfo {author} {\bibfnamefont {N.}~\bibnamefont {Read}},\ }\href
  {\doibase 10.1103/PhysRevB.47.7312} {\bibfield  {journal} {\bibinfo
  {journal} {Phys. Rev. B}\ }\textbf {\bibinfo {volume} {47}},\ \bibinfo
  {pages} {7312} (\bibinfo {year} {1993})}\BibitemShut {NoStop}%
\bibitem [{\citenamefont {Zhang}\ \emph {et~al.}(1989)\citenamefont {Zhang},
  \citenamefont {Hansson},\ and\ \citenamefont {Kivelson}}]{zhk1989}%
  \BibitemOpen
  \bibfield  {author} {\bibinfo {author} {\bibfnamefont {S.~C.}\ \bibnamefont
  {Zhang}}, \bibinfo {author} {\bibfnamefont {T.~H.}\ \bibnamefont {Hansson}},
  \ and\ \bibinfo {author} {\bibfnamefont {S.}~\bibnamefont {Kivelson}},\
  }\href {\doibase 10.1103/PhysRevLett.62.82} {\bibfield  {journal} {\bibinfo
  {journal} {Phys. Rev. Lett.}\ }\textbf {\bibinfo {volume} {62}},\ \bibinfo
  {pages} {82} (\bibinfo {year} {1989})}\BibitemShut {NoStop}%
\bibitem [{\citenamefont {Son}(2015)}]{son2015}%
  \BibitemOpen
  \bibfield  {author} {\bibinfo {author} {\bibfnamefont {D.~T.}\ \bibnamefont
  {Son}},\ }\href {\doibase 10.1103/PhysRevX.5.031027} {\bibfield  {journal}
  {\bibinfo  {journal} {Phys. Rev. X}\ }\textbf {\bibinfo {volume} {5}},\
  \bibinfo {pages} {031027} (\bibinfo {year} {2015})}\BibitemShut {NoStop}%
\bibitem [{\citenamefont {Kane}\ \emph {et~al.}(2002)\citenamefont {Kane},
  \citenamefont {Mukhopadhyay},\ and\ \citenamefont {Lubensky}}]{kml2002}%
  \BibitemOpen
  \bibfield  {author} {\bibinfo {author} {\bibfnamefont {C.~L.}\ \bibnamefont
  {Kane}}, \bibinfo {author} {\bibfnamefont {R.}~\bibnamefont {Mukhopadhyay}},
  \ and\ \bibinfo {author} {\bibfnamefont {T.~C.}\ \bibnamefont {Lubensky}},\
  }\href {\doibase 10.1103/PhysRevLett.88.036401} {\bibfield  {journal}
  {\bibinfo  {journal} {Phys. Rev. Lett.}\ }\textbf {\bibinfo {volume} {88}},\
  \bibinfo {pages} {036401} (\bibinfo {year} {2002})}\BibitemShut {NoStop}%
\bibitem [{Note2()}]{Note2}%
  \BibitemOpen
  \bibinfo {note} {See Supplemental Material.}\BibitemShut {Stop}%
\bibitem [{\citenamefont {Kosterlitz}\ and\ \citenamefont
  {Thouless}(1973)}]{kt1973}%
  \BibitemOpen
  \bibfield  {author} {\bibinfo {author} {\bibfnamefont {J.~M.}\ \bibnamefont
  {Kosterlitz}}\ and\ \bibinfo {author} {\bibfnamefont {D.~J.}\ \bibnamefont
  {Thouless}},\ }\href {\doibase 10.1088/0022-3719/6/7/010} {\bibfield
  {journal} {\bibinfo  {journal} {Journal of Physics C: Solid State Physics}\
  }\textbf {\bibinfo {volume} {6}},\ \bibinfo {pages} {1181} (\bibinfo {year}
  {1973})}\BibitemShut {NoStop}%
\bibitem [{\citenamefont {Moore}\ and\ \citenamefont
  {Read}(1991)}]{mooreread1991}%
  \BibitemOpen
  \bibfield  {author} {\bibinfo {author} {\bibfnamefont {G.}~\bibnamefont
  {Moore}}\ and\ \bibinfo {author} {\bibfnamefont {N.}~\bibnamefont {Read}},\
  }\href {\doibase 10.1016/0550-3213(91)90407-O} {\bibfield  {journal}
  {\bibinfo  {journal} {Nuclear Physics B}\ }\textbf {\bibinfo {volume}
  {360}},\ \bibinfo {pages} {362 } (\bibinfo {year} {1991})}\BibitemShut
  {NoStop}%
\bibitem [{\citenamefont {Haldane}\ and\ \citenamefont
  {Rezayi}(1985)}]{haldane1984}%
  \BibitemOpen
  \bibfield  {author} {\bibinfo {author} {\bibfnamefont {F.~D.~M.}\
  \bibnamefont {Haldane}}\ and\ \bibinfo {author} {\bibfnamefont {E.~H.}\
  \bibnamefont {Rezayi}},\ }\href {\doibase 10.1103/PhysRevB.31.2529}
  {\bibfield  {journal} {\bibinfo  {journal} {Phys. Rev. B}\ }\textbf {\bibinfo
  {volume} {31}},\ \bibinfo {pages} {2529} (\bibinfo {year}
  {1985})}\BibitemShut {NoStop}%
\bibitem [{\citenamefont {Gradshteyn}\ and\ \citenamefont
  {Ryzhik}(1980)}]{gradshteyn1980}%
  \BibitemOpen
  \bibfield  {author} {\bibinfo {author} {\bibfnamefont {I.~S.}\ \bibnamefont
  {Gradshteyn}}\ and\ \bibinfo {author} {\bibfnamefont {I.~M.}\ \bibnamefont
  {Ryzhik}},\ }\href@noop {} {\emph {\bibinfo {title} {Table of integrals,
  series, and products}}}\ (\bibinfo  {publisher} {Academic Press, New York},\
  \bibinfo {year} {1980})\ p.\ \bibinfo {pages} {921}\BibitemShut {NoStop}%
\bibitem [{\citenamefont {Verlinde}\ and\ \citenamefont
  {Verlinde}(1987)}]{verlinde1986}%
  \BibitemOpen
  \bibfield  {author} {\bibinfo {author} {\bibfnamefont {E.}~\bibnamefont
  {Verlinde}}\ and\ \bibinfo {author} {\bibfnamefont {H.}~\bibnamefont
  {Verlinde}},\ }\href {\doibase 10.1016/0550-3213(87)90219-7} {\bibfield
  {journal} {\bibinfo  {journal} {Nuclear Physics B}\ }\textbf {\bibinfo
  {volume} {288}},\ \bibinfo {pages} {357 } (\bibinfo {year}
  {1987})}\BibitemShut {NoStop}%
  \bibitem [{\citenamefont {Arovas}\ and\ \citenamefont
  {Girvin}(1992)}]{arovas1992}%
  \BibitemOpen
  \bibfield  {author} {\bibinfo {author} {\bibfnamefont {D.~P.}\ \bibnamefont
  {Arovas}}\ and\ \bibinfo {author} {\bibfnamefont {S.~M.}\ \bibnamefont
  {Girvin}},\ }in\ \href {\doibase 10.1007/978-1-4615-3466-2_21} {\emph
  {\bibinfo {booktitle} {Recent Progress in Many-Body Theories}}},\
  Vol.~\bibinfo {volume} {3},\ \bibinfo {editor} {edited by\ \bibinfo {editor}
  {\bibfnamefont {T.~L.}\ \bibnamefont {Ainsworth}}, \bibinfo {editor}
  {\bibfnamefont {C.~E.}\ \bibnamefont {Campbel}}, \bibinfo {editor}
  {\bibfnamefont {B.~E.}\ \bibnamefont {Clements}}, \ and\ \bibinfo {editor}
  {\bibfnamefont {E.}~\bibnamefont {Krotscheck}}}\ (\bibinfo  {publisher}
  {Plenum Press},\ \bibinfo {address} {New York},\ \bibinfo {year} {1992})\
  pp.\ \bibinfo {pages} {315--344}\BibitemShut {NoStop}%
\bibitem [{\citenamefont {Trugman}\ and\ \citenamefont
  {Kivelson}(1985)}]{trugman1985}%
  \BibitemOpen
  \bibfield  {author} {\bibinfo {author} {\bibfnamefont {S.~A.}\ \bibnamefont
  {Trugman}}\ and\ \bibinfo {author} {\bibfnamefont {S.}~\bibnamefont
  {Kivelson}},\ }\href {\doibase 10.1103/PhysRevB.31.5280} {\bibfield
  {journal} {\bibinfo  {journal} {Phys. Rev. B}\ }\textbf {\bibinfo {volume}
  {31}},\ \bibinfo {pages} {5280} (\bibinfo {year} {1985})}\BibitemShut
  {NoStop}%
\bibitem [{\citenamefont {Kane}\ \emph {et~al.}(1991)\citenamefont {Kane},
  \citenamefont {Kivelson}, \citenamefont {Lee},\ and\ \citenamefont
  {Zhang}}]{kane1991}%
  \BibitemOpen
  \bibfield  {author} {\bibinfo {author} {\bibfnamefont {C.~L.}\ \bibnamefont
  {Kane}}, \bibinfo {author} {\bibfnamefont {S.}~\bibnamefont {Kivelson}},
  \bibinfo {author} {\bibfnamefont {D.~H.}\ \bibnamefont {Lee}}, \ and\
  \bibinfo {author} {\bibfnamefont {S.~C.}\ \bibnamefont {Zhang}},\ }\href
  {\doibase 10.1103/PhysRevB.43.3255} {\bibfield  {journal} {\bibinfo
  {journal} {Phys. Rev. B}\ }\textbf {\bibinfo {volume} {43}},\ \bibinfo
  {pages} {3255} (\bibinfo {year} {1991})}\BibitemShut {NoStop}%
\bibitem [{Note3()}]{Note3}%
  \BibitemOpen
  \bibinfo {note} {In Supplemental Section II\cite {Note2} we also introduce a
  second set of operators $P_{e,h}(z)$ that annihilate $|\Psi _m\rangle $ and
  define a second term in ${\protect \cal H}_m$ that can also contribute to the
  energy gap.}\BibitemShut {Stop}%
\end{thebibliography}

\begin{thebibliography}{10}

\bibitem{kml2002_SM} C. L. Kane, R. Mukhopadhyay, and T. C. Lubensky, Phys. Rev. Lett. {\bf 88}, 036401 (2002).
\bibitem{kane1991_SM} C. L. Kane, S. Kivelson, D. H. Lee, and S. C. Zhang, Phys. Rev. B {\bf 43}, 3255 (1991).
\bibitem{trugman1985_SM} S. A. Trugman and S. Kivelson, Phys. Rev. B {\bf 31}, 5280 (1985).

\end{thebibliography}

\end{document}